\documentstyle[prd,aps,epsf]{revtex}

\draft

\begin{document}
\thispagestyle{empty}
{\baselineskip0pt
\leftline{\large\baselineskip16pt\sl\vbox to0pt{
\vss}}
\rightline{\baselineskip16pt\rm\vbox to20pt{
\hbox{YITP-00-33}
\hbox{WU-AP/106/00}
\hbox{OU-TAP 140}
\vss}}
}


\vskip1cm
  
\begin{center}{\large \bf
Birth of timelike naked singularity
 }
\end{center}

\vskip1cm

\begin{center}
{\large Hideaki Kudoh \footnote{ Electronic address:
  kudoh@yukawa.kyoto-u.ac.jp}

{\em Yukawa Institute for Theoretical Physics, Kyoto University, Kyoto
606-8502, Japan}}
\\
\vskip0.3cm
{\large Tomohiro Harada \footnote{Electronic
 address:harada@gravity.phys.waseda.ac.jp}

{\em Department of Physics, Waseda University, Shinjuku, Tokyo
169-8555, Japan}}
\\
\vskip0.3cm
{\large Hideo Iguchi \footnote{Electronic
 address:iguchi@vega.ess.sci.osaka-u.ac.jp}

{\em
Department of Earth and Space Science, Graduate School of Science,
     Osaka University, Toyonaka, Osaka 560-0043, Japan
}}

\end{center}

\begin{abstract}
We investigate the causal structure of the  
Harada-Iguchi-Nakao (HIN)'s exact
solution in detail, which  describes the dynamical formation of naked
singularity in the collapse of a regular spherical cluster of 
counterrotating particles. 
There are three kinds of radial null geodesics in the HIN
spacetime.  One is the regular null geodesics and the other two are
the  null geodesics which terminate at the singularity.
The central massless singularity is timelike naked
singularity and 
satisfies the strong curvature condition along the null 
geodesics except for the instant of singularity formation.
The cluster dynamically asymptotes to the singular static 
Einstein cluster in which centrifugal force is balanced 
with gravity. 
The HIN solution provides an interesting example which demonstrates that
collisionless particles invoke timelike naked singularity.
\end{abstract}

\section{Introduction}

 The  cosmic censorship hypothesis (CCH) is one of the most important
 open problems in general relativity since it plays an important role in 
 theories of black hole physics \cite{penrose1979,he1973}.
 The CCH roughly states that
singularities forming in gravitational collapse must be hidden behind event horizons and hence invisible to outside observers.
Many types of gravitational collapse have been studied so far in the
 context of the CCH. Some of them produce globally naked
 singularities, but they cannot
immediately be counterexamples to the CCH because the CCH requires
 suitable matter and appropriate initial conditions. 

 The well-known example of spacetime which describes the  dynamical
formation  of naked singularity  is the Lema\^{\i}tre-Tolman-Bondi (LTB)
  spacetime \cite{tolman1934,bondi1947}.
The LTB spacetimes describe the gravitational collapse of 
a spherically symmetric dust
cloud and have an explicit form of the entire spacetime metric.
It has been proved that the LTB spacetimes have ingoing  null naked
singularity from the generic initial data~\cite{es1979,christodoulou1984,newman1986,jd1993,sj1996,jjs1996}.
However this model is rather simplified because pressure is not taken
into account. 

Some aspects of the effect of pressure on the naked singularity 
formation have been investigated.
Several important results have been obtained.
Ori and Piran investigated the spherically symmetric 
collapse of a perfect fluid
numerically under the assumption of self-similarity
\cite{op1987-1988-1990}, and analytic discussions based on
self-similarity followed it \cite{jd1992-1993}. 
Harada also solved numerically the  non-self-similar
spherically symmetric collapse of a perfect fluid \cite{harada1998}.
Their results are summarized as that naked singularity
can occur in the spherically symmetric collapse of a
perfect fluid for a sufficiently soft equation of state. 
In contrast to isotropic pressure, 
to include only tangential pressure is turned out to be
more tractable \cite{sw1997,magli1997a}. Among the
solutions with vanishing radial pressure, there is a system of a  
spherical cloud of counterrotating particles, in which the 
physical origin of
tangential pressure is clear 
\cite{einstein1939,datta1970,bondi1971,evans1976}.
The each particle in the cluster has its angular 
momentum so that the  average effect of all particles is a non-vanishing tangential pressure. 
 This model is particularly interesting from a physical point of
view because it gives insights into rotational effects on the
gravitational collapse, without raising terrible  difficulties. Harada,
Iguchi and Nakao (HIN) have recently analyzed  singularity 
occurrence in this model  \cite{hin1998,JhinganMagli1999}.
Using the mass-area 
coordinates, which were first introduced by Ori \cite{oli1990} and were applied in this system by Magli \cite{magli1998}, HIN found a new exact  solution
that describes dynamical formation of massless naked singularity. The
results show that the counter-rotation  undresses
the covered singularity, and in particular tangential pressure and
rotation may induce the formation of 
naked singularity. Since the HIN solution is given in the
mass-area coordinates, the motion of each shell and the global
properties of the spacetime are not trivial. We study in this paper
the HIN solution in detail to make them clear.

 The plan of this paper is as follows. In the next section, we
 review the HIN solution.
In Sec. \ref{sec:nullgeodesics}, we study null geodesics
in the HIN spacetime 
which are necessary to determine the causal structure of spacetime.
It is discussed in
Sec. \ref{sec:causalstructure} 
that the central singularity is timelike.
In Sec. \ref{sec:asymptotic},
we will see that the collapse asymptotes to the singular static 
Einstein cluster.
Section \ref{sec:conclusions} is devoted to conclusions.
We use units with $c=G=1$ and follow the sign conventions of the
textbook by Misner, Thorne and Wheeler  about the metric, Riemann and
Einstein tensors \cite{mtw1979}.

\section{The HIN spacetime}
\label{sec:HINsol}
\subsection{The HIN solution}

The HIN solution is the spherical cloud of counterrotating
particles which is marginally bound.  The specific angular
momentum $L(r)$ of each particle at comoving radius $r$ equals to $4F(r)$
\cite{hin1998}, where $F(r)$ is the conserved Misner-Sharp mass function \cite{ms1964}.
Using comoving coordinates, the line element for this system is
reduced to

\begin{equation}
ds^2 = -e^{2\nu(t,r)}dt^2+R^{\prime}{}^{2} (t,r) \left(
1+\frac{16F^2(r)}{R^2(t,r)} \right)dr^2+R^2(t,r)(d\theta^2 +
\sin^2\theta d\phi^2),
\label{eq:ds}
\end{equation}
where $\nu (t,r)$ and $R(t,r)$ satisfy the following set of coupled
partial derivative equations:

\begin{eqnarray}
\nu^{\prime}(t,r) &=& \frac{16F^2}{R(R^2+16F^2)}R '(t,r),
\label{eq:comlapse}
\\
V &\equiv& e^{- \nu} \dot{R}(t,r) = - \frac{|R-4F|
\sqrt{2F}}{\sqrt{R(R^2+16F^2)}}.
\label{eq:comenergy}
\end{eqnarray}
The prime and overdot denote the partial derivatives with respect to $r$ and
$t$, respectively. The energy density $\epsilon(t,r)\equiv-T^t_t$ and
the tangential pressure $\Pi(t,r)\equiv 
T^{\theta}_{\theta}=T^{\phi}_{\phi}$ are given by 

\begin{eqnarray}
\epsilon(t,r) &=& \frac{F^{\prime}}{4\pi R^2 R^{\prime}},
\label{eq:density}
\\
\Pi(t,r) &=& \frac{1}{2} \frac{16F^2}{R^2+16F^2} \epsilon.
\end{eqnarray}
Using the coordinate transformation  $m=F(r)$ and $R=R(t,r)$, 
the HIN solution is given in the mass-area coordinates $(m,R)$ by 

\begin{equation}
ds^2= -A dm^2- 2B dRdm -C dR^2 +R^2 (d\theta^2 +\sin^2\theta
d\phi^2),
\label{eq:ds-mR}
\end{equation}
where $A$, $B$ and $C$ are given by
\begin{eqnarray}
A&=&H \left(1-\frac{2m}{R} \right), \label{eq:A}
\\
B&=&-\frac{R}{|R-4m|}\sqrt{\frac{RH}{2m}},   \label{eq:B}
\\
C&=& \frac{1}{V^2} =\frac{R(R^2+16m^2)}{2m(R-4m)^2},   \label{eq:C}
\\
\sqrt{H(m,R)}&\equiv& \sqrt{B^2-AC}= \frac{(F^{-1})_{,m}((F^{-1})^2+16m^2)}
{|F^{-1}-4m|\sqrt{2mF^{-1}}} \nonumber 
\\ 
& &{}+\mbox{sign}(F^{-1}-4m)\left[\left( \frac{R^2-16 m
R+144m^2}{3\sqrt{2}m(R-4m)}\sqrt{\frac{R}{m}}
+4\sqrt{2} \ln \frac{\sqrt{R}+2\sqrt{m}}
{|\sqrt{R}-2\sqrt{m}|}\right)\right. \nonumber 
\\ 
& &\left.{}-\left( \frac{(F^{-1})^2-16 m
F^{-1}+144m^2}{3\sqrt{2}m(F^{-1}-4m)} \sqrt{\frac{F^{-1}}{m}}
+4\sqrt{2}\ln\frac{\sqrt{F^{-1}}+2\sqrt{m}}
{|\sqrt{F^{-1}}-2\sqrt{m}|} \right) \right].  
\label{eq:sqrtH}
\end{eqnarray}
We denote the inverse function of $F(r)$ as  $F^{-1}=F^{-1}(m)$ and 
 have set $R=F^{-1}(m)$ on the initial spacelike hypersurface,
which corresponds to $R(0,r)=r$. The energy density in the mass-area
coordinates is given by 

\begin{equation}
\epsilon(R,m) = \frac{\sqrt{R^2+16m^2}}{4\pi R^3 |V|\sqrt{H}}.
\end{equation}

We can assume that the metric functions are $C^{\infty}$ class 
with respect to the local Cartesian coordinates at
least in the neighborhood of the center $r=0$ before encountering a
central singularity. By this assumption, the metric variables in the 
comoving coordinates are expanded near the center as

\begin{eqnarray}
\nu (t,r) &=& \nu _0(t) +\nu _2(t)r^2+ \nu _4(t) r^4+ O(r^6)  , 
\label{eq:expantionNu}
\\
R(t,r)   &=& R_1(t)r + R_3(t) r^3 +R_5(t)r^5 + O(r^7) .
\label{eq:expantionR}
\end{eqnarray}
Then from  Eq. (\ref{eq:comenergy}), the arbitrary
mass function $F(r)$ is also expanded as

\begin{equation}
F(r) = F_3 r^3 +F_5 r^5 +F_7 r^7 + O(r^9), 
\label{eq:expantionF}
\end{equation}
and the inverse function $F^{-1}(m)$ is approximately given by 

\begin{equation}
F^{-1}(m)=r= \left( \frac{m}{F_3} \right)^{1/3}-
\frac{F_5}{3{F_3}^2} m + \frac{4{F_5}^2-3F_3
F_7}{9{F_3}^{11/3}}m^{5/3}+ O(m^{7/3}), 
\label{eq:expantionIF}
\end{equation}
where we expand $F^{-1}(m)$ up to the order of $m^{5/3}$ because 
all terms up to this order 
are needed for later calculations. 
We can set $\nu_0 (t)=0$ by using the rescaling freedom of the time
coordinate,  and  Eq. (\ref{eq:comlapse}) yields  $\nu
_2(t)=0 $ .  Thus the leading order of $\nu(t,r)$ is given
by
\begin{equation}
\nu (t,r) = \frac{4 {F_3}^2}{{R_1}^2(t)} r^4+ O(r^6)  .
\label{eq:BehaviorNu}
\end{equation}
For $R(t,r)$, we obtain 

\begin{equation}
R(t,r) = \left[1-\frac{t}{t_0} \right]^{2/3} r +O(r^3), 
\label{eq:BehaviorR}
\end{equation}
where $t_0$ is given by 
\begin{equation}
t_0 = \frac{1}{3} \sqrt{\frac{2}{F_3}}.
\label{eq:NStime}
\end{equation}
The break down of these expansions means the
appearance of central singularity. Since the break down is given by
$R_1(t)=0$,  the singularity formation
time  is $t=t_0$.

On the initial spacelike hypersurface, the regularity requires
$R(t=0,r)=r > 4F= {  O}(r^3)$ in a sufficiently small, but finite region
around $r=0$.  From Eq. (\ref{eq:comenergy}),  $R(t,r)$ is a monotonically 
decreasing function with respect to  $t$, and then  each initially collapsing shell approaches $R=4F$. 
Using proper time $\tau(t,r) = \int e^{\nu}dt$ of each shell, 
the behavior of this approach is 
\begin{equation}
R-4F \propto \exp \left(- \frac{\tau}{8F} \right).
\label{eq:R-4F}
\end{equation}
This behavior shows that $R$ approaches $4F$ asymptotically and thus
$R$ is always $R \geq 4F$ if it initially holds. Since the trapped
region is given by $0 \leq R<2F$,  the region around the center is not
trapped eternally. Hence  the central singularity is globally naked.

\subsection{Algebraic root equation method}

The nakedness of the central singularity in the HIN spacetime is also 
confirmed by examining the algebraic root equation, which probes the
existence of outgoing radial  null geodesics from the center
\cite{jd1993,sj1996,jjs1996,magli1998}. To obtain the root equation,
consider the radial null geodesic equation in mass-area coordinates
\begin{equation}
\frac{dR}{dm}=\frac{-B \mp \sqrt{H} }{C}= J_{\mp}\sqrt{H(m,R)} \, ,
\label{eq:nulleq}
\end{equation}
where $J_{\mp}$ is given by  
\begin{equation}
J_{\mp} \equiv |V|\left( \frac{R}{\sqrt{R^2+16F^2}} \mp |V| \right) =
\frac{|z-1|}{(z^2+1)\sqrt{2z}} \left(z \mp \frac{|z-1|}{\sqrt{2z}}
\right),
\label{eq:Jmp}
\end{equation}
and we have defined $z\equiv R/4m(\geq 1)$ for later convenience. The upper
and lower signs refer to outgoing and ingoing null geodesics in the
collapse phase, respectively. Hereafter  we use this sign
convention. To investigate the behavior of null geodesics near the
center, we define 
\begin{equation}
x(m)= \frac{R}{2m^{\alpha} },
\label{eq:x(m)}
\end{equation}
where $\alpha$  is a constant.
Applying l'Hospital theorem to  
Eq. (\ref{eq:x(m)}), we obtain the root equation

\begin{eqnarray}
x_0 &=& \lim _{m \to 0 } \frac{m^{1-\alpha}}{2\alpha} \frac{dR}{dm}
\nonumber
\\
&=& \lim _{m \to 0 } \frac{ m^{3(1-\alpha)/2}( x_0 -2m^{1-\alpha})}{2
\alpha \, x_0\left({x_0}^2+4m^{2(1-\alpha)} \right)} 
\, \sqrt{H(m,2  x_0\,m^{\alpha})} \left[ x_0 ^{3/2} \mp m^{(1-\alpha)/2} (x_0 -2m^{1-\alpha}) \right],
\label{eq:rootEq}
\end{eqnarray}
where $x_0 \equiv x(0)$ is introduced.
If we find a consistent set of  $x_0$ and $\alpha$, 
we have  null geodesics which behave as Eq. (\ref{eq:x(m)}) 
terminating at the center $m=0$.  The root equation method picks up
only the  geodesics behaving as Eq. (\ref{eq:x(m)}). To find the
another possible null geodesics, we must solve the null geodesic equation.

\section{Null geodesics in the HIN spacetime}
\label{sec:nullgeodesics}
From the detailed analysis of Eq.(\ref{eq:rootEq}), we can prove that
all possible values of $\alpha$ for positive finite $x_{0}$ 
are 1/3, 7/9 and 1.
In reality, these three values of $\alpha$ correspond
to three different types of null geodesics in the HIN spacetime, and
it will be proved in this section that there are no other null
geodesics in the HIN spacetime.
We will see below the derivations and properties of 
these three types of null geodesics in detail.

\subsection{$\alpha=1/3$ : regular null geodesics}
The null geodesics with $\alpha=1/3$ correspond to 
regular null geodesics. 
For this case, the value of 
$x_0$ for $\alpha=1/3$ cannot be determined 
by the root equation method. 
We will see that the 
regular null geodesics actually have $\alpha=1/3$.
From Eqs. (\ref{eq:ds}), (\ref{eq:BehaviorNu}) and
(\ref{eq:BehaviorR}), the radial null geodesic equation for lowest order is 
\begin{equation}
\frac{dt}{dr} \approx \pm \left[ 1-\frac{t}{t_{0}} \right]^{2/3}.
\label{eq:NullEq-dtdr}
\end{equation}
Inserting the solution $t=t(r)$ of Eq. (\ref{eq:NullEq-dtdr}) to
Eq. (\ref{eq:BehaviorR}), we obtain  the regular null
geodesics in the mass-area coordinates
\begin{equation}
R(m) \approx \left( 1- \frac{t(0)}{t_{0}} \right)^{2/3} 
\left(\frac{m}{F_3}  \right)^{1/3},
\label{eq:NullgeodesicRm}
\end{equation}
where $t(0)<t_{0}$ is the time when the null geodesics arrive at or emanate from the regular center $r=0$.  The arbitrariness of $x_0$ for
$\alpha=1/3$ comes from the arbitrary
constant $t(0)$ of Eq. (\ref{eq:NullgeodesicRm}). 

We can find that the scalar curvature is finite at the center
along these null geodesics (see Appendix \ref{appen:ScalarCurvature}).
In fact, we can prove that all the null geodesics with $\alpha=1/3$
terminate at or emanate from the regular center.
The  integration of  Eq. (\ref{eq:comenergy}) by using the proper time
$\tau(t,r)$ is 
\begin{equation}
\tau(t,r) =  c(m) - 4m \int ^{z} _{R_0 (m)/4m}
\frac{\sqrt{2s(s^2+1)}}{s-1} ds,
\label{eq:tau}
\end{equation}
where $R_0(m)$ is the initial area radius $R(0,r)=F^{-1}(m)$ and $c(m)$ 
is an arbitrary function of $m$.  For the regular spacetime,
$\nu(t,0)=\nu_{0}(t)=0$ implies that we can set $\tau=t$ on
$m=0$. Then  the singularity appears on $m=0$ when the proper time is
$\tau = t_0$. The elliptic integral of Eq. (\ref{eq:tau}) is
explicitly  integrated by approximating the integrand.  $R_0(m)/4m \gg 1$ is 
satisfied around the regular center $m=0$  and if we consider  the
condition  $z \gg 1$, the integrand of Eq. (\ref{eq:tau}) is approximately
$\sqrt{2s}$. Then, $\tau$ for each shell becomes
\begin{equation}
\tau \approx c (m) + t_0 -\frac{\sqrt{2}}{3} 
\left( \frac{R}{m^{1/3}} \right) ^{3/2}.
\label{eq:tau-z>1}
\end{equation}
 We consider the null geodesics of Eq. (\ref{eq:x(m)}) for
$\alpha={1/3}$. From Eq. (\ref{eq:tau-z>1}), the proper time along these geodesics is 
\begin{equation}
\tau = c(m)+ t_0 -\frac{4}{3} x^{3/2}.
\label{eq:tau<tau0}
\end{equation}
For  the regular shell motion of Eq. (\ref{eq:BehaviorR}),  Eq. (\ref{eq:tau-z>1}) becomes
\begin{equation}
\tau = c(m)+t.
\end{equation}
Taking into account $\tau= t $ on $m=0$, we  have $c (0)=0$ and thus
Eq. (\ref{eq:tau<tau0}) implies  $\tau < t _0$ at $m=0$.  Therefore  all the null geodesics behaving as $R\propto m^{1/3}$ terminate at $m=0$ before the
singularity appears.  

\subsection{$\alpha=7/9$ : the earliest singular null geodesic}

The singular null geodesic with $\alpha=7/9$ for both ingoing
and outgoing geodesics was found by Harada, Iguchi and
Nakao \cite{hin1998}. The behavior of the  null geodesic as 
\begin{equation}
R \approx 2x_0 m^{7/9},
\label{eq:R2x0}
\end{equation}
is not regular and  thus it  terminates at the central singularity. 
The coefficient $x_0$ is given by
\begin{equation}
x_0=\left(\frac{24F_3^2-F_5}{4\sqrt{2} F_3^{13/6}}\right)^{2/3},
\end{equation}
for $F_5 <24 {F_3}^2$. Note that $F_5<24{F_3}^2$ is the same as the
requirement of no shell-crossing singularity and this  condition holds
if $\epsilon (0,r)$ is a non-increasing function of $r$.
It is found that Eq. (\ref{eq:tau-z>1}) reduces to $\tau = t _0 $ at
$m=0$ for  $ \frac{1}{3} < \alpha <1$. Thus all the null geodesics
of  $\frac{1}{3}< \alpha < 1 $,  particularly $\alpha =7/9$,
terminate at the singularity. The scalar curvature diverges along the
null geodesic with $\alpha=7/9$ (see Appendix \ref{appen:ScalarCurvature}).

We will see below that there is only one  ingoing or outgoing null
geodesic with $\alpha=7/9$ and that this null geodesic is the first one  which arrives at or emanates from the singularity at $t=t_0$. 
The null geodesic equation in mass-area coordinates has a singular
point at $(m,R)=(0,0)$. To make the singular point tractable, 
we introduce new coordinates as 

\begin{eqnarray}
\chi  &\equiv& m^{1/9},
\label{eq:define-u}
\\
\vartheta &\equiv& \left( \frac{R}{m^{7/9}} \right)^{3/2},
\end{eqnarray}
and then Eq. (\ref{eq:nulleq}) becomes

\begin{equation}
\frac{ d \vartheta}{d\chi }+\frac{6}{\chi }(\vartheta-\lambda)=\lambda
\Psi_{\mp}( \chi , {\vartheta}),
\label{eq:RegularNulleq}
\end{equation}
where $\Psi_{\mp}$ is given by

\begin{eqnarray}
\Psi_\mp(\chi ,\vartheta) &=& \frac{1}{\chi }\left( \frac{\psi_{\mp}(\chi ,\vartheta)}{\lambda}-6 \right),
\label{eq:Phi-mp}
\\
\psi_\mp(\chi ,{\vartheta})&=& \frac{9}{2} \left(3\chi ^2 \vartheta ^{1/3}
\sqrt{H(\chi ,{\vartheta})} J_{\mp}(\chi ,{\vartheta}) -\vartheta \right),
\label{eq:phi-mp}
\end{eqnarray}
and we have  introduced a parameter  $0<  \lambda < \infty $.

The form of Eq. (\ref{eq:RegularNulleq}) is similar to the
form of the null geodesic equation of the  LTB 
spacetime given in~\cite{christodoulou1984,newman1986}.  One may now
follow to Christodoulou's argument to show the existence and
uniqueness of a continuous solution of 
Eq. (\ref{eq:RegularNulleq}) \cite{christodoulou1984}.
It is sufficient to consider the null geodesic equation in the   
neighborhood of the center $\chi =0$ since we are interested in the     
radial null geodesics terminating at $\chi =0$.  There is no 
singular point at $\chi >0$ where $\vartheta$ is strictly positive by
 definition. Problems appear  when we consider the center $\chi =0$.
We expand $J_{\mp}$ and $\sqrt{H}$ around $\chi =0$  using Eq. (\ref{eq:expantionIF}),

\begin{eqnarray}
J_{\mp} &\approx& \sqrt{2}\left(\frac{\chi }{\vartheta^{1/3}}\right) 
\mp 2\left( \frac{\chi }{\vartheta ^{1/3}} \right)^2 
- 4\sqrt{2} \left(\frac{\chi }{\vartheta^{1/3}} \right)^{3}
+{ O} \left( \left( \frac{\chi }{\vartheta^{1/3}} \right)^{4} \right),
\label{eq:JmpExpantion-z}
\\
\sqrt{H} &\approx& \frac{8 {x_{0}}^{3/2}}{9} \frac{1}{\chi ^3}
+\frac{1}{3 \sqrt{2}} \left(\frac{\vartheta ^{1/3}}{\chi} \right)^{3} 
- 2\sqrt{2} \left( \frac{\vartheta ^{1/3}}{\chi} \right)+ {O}(\chi
^3)+{O} \left( \frac{\chi}{\vartheta^{1/3}} \right), 
\label{eq:sHExpantion-z}
\end{eqnarray}
and from this expansion, $ \psi_{\mp}$ is 

\begin{equation}
\psi_{\mp}(\chi ,{\vartheta}) \approx \psi_0+\psi_1(\vartheta) \chi  +\psi_2(\vartheta)
\chi ^2+{  O}(\chi ^3),
\end{equation}
where the coefficients of each order are
\begin{eqnarray}
\psi_0 &=& 12 \sqrt{2} \, x_0 ^{3/2} , 
\nonumber
\\
\psi_1(\vartheta) &=& \mp \frac{ \, \vartheta^{2/3}}{\sqrt{2}} \left( 9+ \frac{2\psi_0}{\vartheta} \right) ,
\\
\psi_2 (\vartheta) &=& -12 \vartheta ^{1/3}\left(6+\frac{ \psi_0}{3 \vartheta} \right).
\nonumber
\end{eqnarray}
If  we choose the  parameter $\lambda$ to be $\lambda =\lambda_0 \equiv
 (2x_0)^{3/2}$,  $\Psi_{\mp}$ is at
least $C^1$ in $\chi  \geq 0, \vartheta>0 $. 
We can apply the contraction mapping principle to
Eq. (\ref{eq:RegularNulleq}) to find that there exists the 
solution satisfying  $\vartheta(0)=\lambda _0$,  and moreover that it is
the unique solution to Eq. (\ref{eq:RegularNulleq}) which is
 continuous at $\chi=0$.  This solution with $\vartheta(0)=\lambda_0$ exactly agrees with the geodesics of Eq. (\ref{eq:R2x0}).
The proof is presented in Appendix \ref{appen:uniqueness}.
Therefore there is no other solution with  $0< \vartheta(0) 
< \infty $. In other words, another possible solution must be
$\vartheta(0)=0$ or $\infty$.

We consider possible solutions with $\vartheta(0)=\infty$. When
$\vartheta(0)=\infty$,  Eq. (\ref{eq:RegularNulleq}) is approximated around $\chi =0$ by
using Eqs. (\ref{eq:JmpExpantion-z}) and (\ref{eq:sHExpantion-z}) which are valid even in this limit as follows:
\begin{equation}
\frac{d\vartheta}{d \chi } \approx - \frac{6\vartheta}{\chi } .
\label{eq:regular-drdu}
\end{equation}
The integration of  this equation gives  $\vartheta(\chi ) \propto
1/{\chi}^6$. 
This behavior coincides with the regular null
geodesics of Eq. (\ref{eq:NullgeodesicRm})
 because all the geodesics behaving as $R \propto m^{1/3}$ terminate at
the regular center as we have shown before.

It is important that $R(t,r)$ is a monotonically decreasing function
with  respect to $t$, and thus that $\vartheta$  decreases as $t$ increases
when $\chi$ is fixed.  Because  of this time direction in the mass-area
coordinates and the fact that there are no other null geodesics with
$(2x_0)^{3/2}< \vartheta(0)< \infty$, the geodesic with $\vartheta(0) =(2x_0)^{3/2}$ 
is  the first null geodesic which arrives at or  emanates  from  the
appeared singularity.  Hence we  conclude  that the  arrival or  emanational time in comoving coordinates is the singularity formation time $t=t_0$.

\subsection{$\alpha=1$ : later singular null geodesics}

There are  also null geodesics with $\alpha=1$.  $x(m)$ is expressed for
$\alpha=1$ by a non-analytic function
\begin{equation}
x(m) \approx 2 +2 \exp \left(- \frac{D}{ \sqrt{m}} \right),
\end{equation}
where $D$ is the positive constant which parameterizes the null
geodesics. We find that the scalar curvature diverges at the
center along these null geodesics as is seen
in Appendix \ref{appen:ScalarCurvature}.
We will see these null geodesics in detail below.

In the coordinate $\vartheta$, these null geodesics are
described by solutions with $\vartheta(0)=0$ if they exist. 
We first search the solution with
$z(0)=\infty$ and  $\vartheta(0)=0$. Under these conditions,
Eq. (\ref{eq:nulleq}) reduces around $m=0$ to  
\begin{equation}
\frac{dz}{dm}=\frac{1}{4m}\left( \sqrt{H}\,J_{\mp} -4z\right)
\approx 
\frac{1}{m\sqrt{z}}\left
( \frac{\sqrt{2}{x_0}^{3/2}}{9}m^{-1/3}-\frac{2}{3}z^{3/2} \right).
\label{eq:asymNullEq}
\end{equation}
For $z^{3/2} \ll  m^{-1/3}$, we can neglect the second term of
Eq. (\ref{eq:asymNullEq}) and immediately  integrate it. However, the
solution contradicts  $z^{3/2} \ll  m^{-1/3}$ because it is given by
$z \propto m^{-2/9}$. For $z^{3/2} \propto m^{-1/3}$ or  $z^{3/2} \gg
m^{-1/3}$, there are consistent solutions. 
The solution of  Eq. (\ref{eq:asymNullEq}) for $z^{3/2} \gg m^{-1/3}$ is  
\begin{equation}
z\propto \frac{1}{m^{2/3}},
\nonumber
\end{equation}
which corresponds regular null geodesics with $\alpha =1/3$.
When  $z^{3/2} \propto  m^{-1/3}$, the integration gives  
\begin{equation}
z \propto m^{-2/9},
\nonumber
\end{equation}
which corresponds to the unique null geodesic with $\alpha
=7/9$. These two solutions satisfy  the condition $z(0)=\infty$, but
do not  satisfy $\vartheta(0)=0$.  Consequently there is no solution which
satisfies these two conditions.  

We consider the geodesics with $z(0)< \infty$ and $\vartheta(0)=0$. Using
l' Hospital theorem, we obtain a restriction on $J_{\mp} \sqrt{H}$ as 
\begin{equation}
4 z(0)= \lim _{m \to 0}\frac{dR}{dm}
= \lim _{m \to 0} J_{\mp} \sqrt{H}
< \infty.
\label{eq:condition-z(0)}
\end{equation}
Though  $J_{\mp} $ is strictly positive at $m=0$ as long as  $z(0)>1$,
$\sqrt{H}$ diverges to positive infinity at $m=0$ as
\begin{equation}
\sqrt{H} = 4\sqrt{2} \left( \frac{ \beta }{ m^{1/3}}+\gamma m^{1/3}+
{  O}(m)\right)
+\frac{4\sqrt{2}}{3}\frac{\sqrt{z}(z^2-4z+9)}{z-1}
+ 4\sqrt{2} \ln  \frac{\sqrt{z}+1}{\sqrt{z}-1},
\label{eq:expantionH-atm0}
\end{equation}
where we have defined
\begin{eqnarray} 
\beta &\equiv& \frac{\sqrt{2}}{9}x_{0}^{3/2} ,\\  
\gamma &\equiv& - \frac{ 1440 {{{F_3}}^4} +
48\,{{{F_3}}^2}\,{F_5} - 17\,{{{F_5}}^2} + 12 {F_3} {F_7}}
{216 {{F_3}}^{23/6}}.
\end{eqnarray} 
Thus, there exists no solution with the boundary condition $1<z(0)<
\infty $. However it may be  possible for null geodesics to satisfy
Eq. (\ref{eq:condition-z(0)}) only if $z(0)=1$.

We  introduce the new coordinate $y\equiv z-1$ to study
the solution $y(m)$ with $y(0)=0$. By expanding Eq. (\ref{eq:nulleq})
around $y=0$, we obtain the consistent solution with $y(0)=0$ in the
lowest order approximation.  We give here the null geodesic 
equation which is expanded up to the enough order
 so that  the difference between ingoing and outgoing null geodesics
appears in the expansion;
\begin{equation}
\frac{dy}{dm} \approx \frac{y}{2m}\left( -\ln y
+\frac{\beta}{m^{1/3}}+ \delta \mp \sqrt{2}
+{\gamma} \, m^{1/3}  \right).
\end{equation}
It can be integrated to
\begin{equation}
y \approx  \exp \left( -\frac{D}{\sqrt{m}}+\frac{3\beta}{m^{1/3}}+ \delta 
\mp \sqrt{2}+\frac{3\gamma}{5} \, m^{1/3}  \right),
\label{eq:solution-y}
\end{equation}
where $D>0$ is an integration constant and $\delta $ is defined as
$\delta \equiv 2 \ln 2 -\frac{8}{3}$.  We will see in
Sec. \ref{sec:asymptotic}  that the constant $D$ is related to
the time when the null geodesics terminate at the center $m=0$.
 The geodesics of Eq. (\ref{eq:solution-y})  satisfy   $y(0)=0$ and
they  are classified  into the geodesics with $\alpha =1$. 
To obtain the proper time when the null  geodesics terminate at
 the singularity, we consider Eq. (\ref{eq:tau}) again.
By the  approximation of the integrand, Eq. (\ref{eq:tau}) reduces
around $m=0$ to

\begin{equation}
   \tau \approx c (m) - 4m \left(2 \ln y(m)- \frac{t _{0}}{4m}\right).
\end{equation}
Then the proper time when the null geodesics terminate at $m=0$ is
$\tau=t _0$   and it  means that the null 
geodesics terminate at the central singularity.
 As a result, it is concluded  that all  the solutions with $\vartheta(0)=0$ are the solutions which terminate at the singularity with
$y(0)=0$.

	\section{Causal structure of the HIN spacetime}
\label{sec:causalstructure}
	\subsection{Timelike singularity}

It is important that Eq. (\ref{eq:solution-y}) 
includes all the null geodesics  emanating from the
center after the singularity appears.
We construct double null coordinates by
using Eq. (\ref{eq:solution-y}) and study the causal structure of the
spacetime which is covered by these  null geodesics.
From Eq. (\ref{eq:solution-y}), we introduce double null coordinates
$(u,v)$ which satisfy around $m=0$
\begin{eqnarray}
2\sqrt{2}u &\approx& \sqrt{m} \left( - \ln y +\frac{3\beta}{m^{1/3}}+ \delta 
- \sqrt{2}+\frac{3\gamma}{5} \, m^{1/3}  \right) ,
\label{eq:defineu}
\\
2\sqrt{2}v &\approx& \sqrt{m} \left( - \ln y +\frac{3\beta}{m^{1/3}}+ \delta 
+ \sqrt{2}+\frac{3\gamma}{5} \, m^{1/3}  \right).
\label{eq:definev}
\end{eqnarray}
Then, $m$ and $y$ are expressed in the null  coordinates by
\begin{eqnarray}
\sqrt{m} &\approx&  v-u,
\label{eq:muv}
\\
y &\approx& \exp \left
( -\frac{\sqrt{2}(v+u)}{v-u}+\frac{3\beta}{(v-u)^{2/3}}+ \delta + 
\frac{3\gamma}{5} \, (v-u)^{2/3}  \right).
\label{eq:yuv}
\end{eqnarray}
We are considering a sufficiently small, 
but finite region around $m=0$. Then 
Eq. (\ref{eq:muv}) restricts $u$ and $v$ to $u \approx v$.
From  Eqs. (\ref{eq:muv}) and (\ref{eq:yuv}), $dm$ and $dR$ are given
by   
\begin{eqnarray}
dm &\approx&2 (v-u)(dv-du),
\label{eq:dm}
\\
dR-4dm 
&\approx& - 4 \sqrt{2} y  \left( v+u -\sqrt{2} \beta (v-u)^{1/12}
+(\sqrt{2}+1)(v-u)+ \frac{\sqrt{2}}{5} \gamma \, (v-u)^{5/12} \right) du 
\nonumber
\\
& &+ 4 \sqrt{2} y  \left( v+u -\sqrt{2} \beta (v-u)^{1/12}
+(\sqrt{2}-1)(v-u)+ \frac{\sqrt{2}}{5} \gamma \, (v-u)^{5/12}
\right) dv .
\label{eq:dR}
\end{eqnarray}
Inserting  Eqs. (\ref{eq:dm}) and (\ref{eq:dR}) into
Eq. (\ref{eq:ds-mR}) and expanding the metric functions $A,B$ and
$C$ around  $(m,y)=(0,0)$, we obtain  the line element in the double null
coordinates as 
\begin{equation}
ds^2 \approx  -512(v-u)^2 dvdu + R^2(d\theta^2 +
\sin^2\theta d\phi^2).
\label{EQ:CPNFORMALLYFLAT}
\end{equation}
The details of the calculation are given in Appendix \ref{appen:Null}.
In the null coordinates, 
the central singularity is represented at $u=v$ which corresponds to 
the center $m=0$ because of Eq. (\ref{eq:muv}). It is found that
the world line of
$u=v$ is timelike. 
Hence the central singularity in the spacetime which is
covered by the null geodesics given 
by Eq. (\ref{eq:solution-y}) is timelike.

\subsection{Penrose diagram}

\begin{figure}
\centerline{\epsfxsize = 5cm \epsfbox{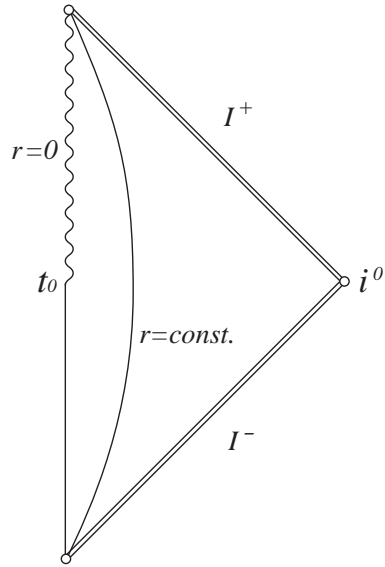}}
\vspace{3mm}
\caption{ The conformal diagram of the HIN spacetime. A timelike naked 
singularity emerges at the center at  $t= t_0$. There are one parameter 
family of both ingoing and outgoing null geodesics which terminate at
and emanate from the naked singularity.}
\protect 
\label{fig:timelikeNS}
\end{figure}

\begin{figure}
\centerline{\epsfxsize = 7cm \epsfbox{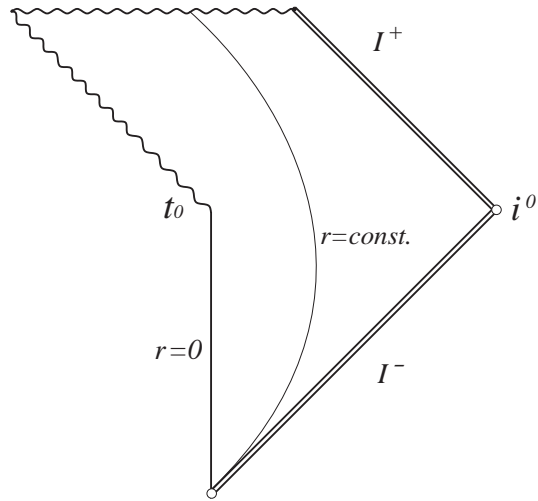}}
\vspace{3mm}
\caption{ The conformal diagram of the LTB spacetime with naked
 singularity. 
There are one parameter family of outgoing null geodesics
which emanate from the naked singularity 
but only one ingoing null geodesic which terminates at
the naked singularity.}
\protect 
\label{fig:LTB}
\end{figure}

Here we summarize the obtained results. The whole of the spacetime is covered by three types 
of null geodesics which can be classified  by the value of $\alpha$
into $\alpha =1/3$, $7/9$ and $1$. The null geodesics with
$\alpha=1/3$ emanate from or  terminate at the regular center and
they are parametrized by one parameter. There is only one null
geodesic with $\alpha=7/9$  and it is the earliest one which emanates
from or terminates  at the naked singularity. The null geodesics with
$\alpha=1$ emanate from or terminate at the timelike singularity and 
they are parametrized by one parameter.   From these results,
it is now possible to draw the conformal diagram of  the HIN spacetime
(see Fig. \ref{fig:timelikeNS}). For comparison, we also present the conformal
diagram of the LTB spacetime  with naked singularity (see Fig. \ref{fig:LTB}).
It is found that the effect of counterrotation makes the singularity timelike in this model.


\subsection{Curvature strength of the singularity}

We should note the curvature strength of the singularity in the HIN
spacetime. According to Tipler and Kr{\'o}lak, 
we classify the curvature strength whether the
singularity satisfies the strong curvature condition (SCC) or the
limiting focusing condition (LFC) \cite{tipler1977,krolak1987}. 
Harada, Nakao and Iguchi studied the SCC and the LFC for spherically
symmetric spacetimes with vanishing radial pressure \cite{hin1999}.
Applying theorems 1, 2 and 3 of their paper to the HIN spacetime,  
we can know the curvature
strength along  each null geodesic.
For $\alpha=7/9$, not the SCC but only the LFC is satisfied 
along the null geodesics. For $\alpha =1$, the SCC is satisfied along the
null geodesics.
In conclusion, the naked singularity is relatively weak 
at the formation  but becomes strong after that.

\section{Asymptotic Behavior}
\label{sec:asymptotic}

It is worth while examining the asymptotic behavior of the 
spacetime. We concern with the null geodesic behavior in the comoving
coordinates  corresponding to  Eq. (\ref{eq:solution-y}).  To obtain
an insight, we consider the asymptotic behavior of the HIN solution in
comoving  coordinates.  Since the time coordinate $t$  at the center 
does not proceed  after the
singularity formation,  we must introduce new  time coordinate  which
no longer agrees with the proper time at $r=0$. We denote the new time
coordinate as $T$ and consider that the time $t$ in Eqs. (\ref{eq:ds}), (\ref{eq:comlapse}) and (\ref{eq:comenergy}) is replaced by $T$.

We take the limit $R = 4F $ because  $R(T,r)$  approaches 
$4F(r)$ asymptotically. From Eq. (\ref{eq:comlapse}), the metric in this
limit is given  by 

\begin{equation}
ds^2 = - \frac{4F(r)}{K(T)}dT^2+2(4F'(r))^2 dr^2 +(4F(r))^2(d\theta^2
+\sin^2\theta  d\phi^2).
\end{equation}
The function $K(T)$ is an arbitrary function of  $T$ which comes from the
integration of Eq. (\ref{eq:comlapse}). For simplicity we set
$K(T)=K= constant$ by rescaling the time coordinate. The energy
density  in this limit is 

\begin{eqnarray}
  \epsilon &=& \frac{1}{256 \pi  F^2}.
\end{eqnarray}
This solution is a static system of counterrotating particles, which
is called the Einstein cluster. In particular, the solution has
timelike naked singularity at the center.

To study the asymptotic behavior of the shell motion
for fixed $r$, we set perturbed quantities $e(T,r)$ and $d(T,r)$ for the
coordinates $(T,r)$ around $R = 4F$ as, 
\begin{eqnarray}
R(T,r)&=& 4F(r) \left[ 1+e(T,r) \right],  
\label{eq:define-e}
\\
e^{2 \nu (T,r)} &=& e^{ d(T,r)}\frac{R(T,r)}{K} \approx
\frac{4F(r)}{K}  [1+e(T,r)+d(T,r)].
\end{eqnarray}
Inserting these quantities into Eqs. (\ref{eq:comlapse}) and
(\ref{eq:comenergy}), we obtain  following perturbed equations in linear order,

\begin{eqnarray}
\dot{e}(T,r) &=& -\frac{e(T,r)}{4 \sqrt{K F(r)}},
\\
d'(T,r)&=& -\frac{F'(r)}{F(r)} e(T,r).
\end{eqnarray}
The solutions of these equations are 

\begin{eqnarray}
e(T,r)  &=& E(r) \exp \left(- \frac{T}{4 \sqrt{K F(r)}} \right),
\label{eq:e(T,r)}
\\
d(T,r)  &=& - \int e(T,r) \frac{F'}{F} dr +G(T),
\label{eq:d(T,r)}
\end{eqnarray}
where $E(r)$ and $G(T)$ are arbitrary functions of the 
comoving radius $r$ and time  $T$ respectively. We have implicitly assumed 
that the time $T$ is very large compared to the singularity formation time.
However, Eqs. (\ref{eq:e(T,r)}) and (\ref{eq:d(T,r)}) imply that this
perturbation scheme is valid soon after the singularity formation time
as long as  we consider the region in which the  radius $r$ is very small.

In the  asymptotic region, the proper time becomes
\begin{equation}
\tau (T,r) = 2\sqrt{\frac{F}{K}}\, \left[\, T + const. + {  O}\left
( e(T,r) \right)\, \right].
\label{eq:tau(T,r)}
\end{equation}
It is found that the asymptotic behavior of Eqs. (\ref{eq:e(T,r)}) and
(\ref{eq:tau(T,r)}) completely coincides with the already 
obtained behavior of Eq. (\ref{eq:R-4F}). Using these results of
perturbation,  the
null geodesic equation in the comoving coordinates is  
\begin{equation}
 \frac{dT}{dr} = \pm \sqrt{\frac{2K}{F}}\Bigl( 2F' + 2Fe'-F'd \Bigr),
\end{equation}
and it is immediately integrated to 

\begin{equation}
T(r) = T(0) \pm \sqrt{2K} \left( 4 \sqrt{F(r)}+ 
\int \frac{dr}{\sqrt{F(r)}}\left[2F e' - F'd\, \right] \right).
\label{eq:t(r)-asy}
\end{equation}
Integration constant $T(0)$ is  the time when these null geodesics
terminate at or emanate from the singularity, and this
parameterization of the null geodesic family corresponds to the fact 
that the singularity is timelike. From Eqs. (\ref{eq:define-e}),
(\ref{eq:e(T,r)}) and  (\ref{eq:t(r)-asy}),  we obtain the null
geodesics in the mass-area coordinates corresponding to
Eq. (\ref{eq:t(r)-asy}), 

\begin{equation}
y= \frac{R}{4F}-1=E(r) \exp \left( -\frac{T(0)}{4\sqrt{KF(r)}} \mp
\sqrt{2} \mp \frac{1}{2 \sqrt{2F(r)}} \int
\frac{dr}{\sqrt{F(r)}}[2Fe'-F'd \,]  \right).
\end{equation}
Comparing this result to Eq. (\ref{eq:solution-y}),
the  parameter $D$  relates to the arrival
time $T(0)$ as $T(0)=4 D \sqrt{K} $.

\section{Conclusions}
\label{sec:conclusions}

We have studied the causal structure of the HIN spacetime and it was
shown that the central massless singularity of this spacetime 
is timelike. To show this fact, we have investigated the radial null
geodesics in detail. The null geodesics are classified to three types,
$\alpha = $ 1/3, 7/9 and 1, by  their dependence of $R$ on  $m$ near the center.  One is regular and the other two are
singular. The classification of the singular geodesics corresponds to  
their arrival or emanational time at the central singularity. 
The $\alpha =7/9$ type is the earliest singular null geodesic which
arrives at or emanates from the singularity at its formation time
$t_0$,  while the geodesics with $\alpha=1$ arrive at or emanate from the
singularity after its appearance.

 We have shown that singular null
geodesics with $\alpha=1$ exactly parametrized by their arrival or emanational time and
that there is only one set of ingoing  and outgoing geodesics
for each parameter. This fact shows that the central singularity has
timelike property. We have also constructed double null coordinates
around the central singularity from the null geodesics with $\alpha=1$.
The line element in this double null coordinates shows that there is
 timelike singularity in this spacetime. 
We have considered the asymptotic behavior of this spacetime 
after the singularity appeared in comoving 
coordinates. It gives us understanding
of the null geodesic behavior in mass-area coordinates, and of 
the parametrization of the null geodesic family.
The curvature strength of this singularity was also investigated.
The LFC is satisfied along the null geodesics with $\alpha =7/9$
and the SCC is satisfied for  $\alpha=1$. The
curvature strength of the naked singularity
is relatively weak at the formation
and it becomes strong after that.

In summary, the HIN solution describes a dynamical
formation of timelike naked singularity, that is, the birth of
timelike singularity.   The solution dynamically
tends to the static singular Einstein cluster as time proceeds.   
Though the HIN system is simply composed of collisionless particles, 
the collapse leads to the nontrivial causal structure.  It implies 
that the effects of rotation and tangential pressure play important
roles in the final stage of collapse, particularly the singularity formation.
 
\acknowledgements

We are grateful to T. Nakamura, H. Kodama, T.P. Singh,
K. Nakao, A. Ishibashi and S.S. Deshingkar for helpful 
discussions. This work was partly supported by the Grant-in-Aid for Scientific
Research (No. 05540) from the Japanese Ministry of Education, Science, 
Sports and Culture.

\appendix

\section{Scalar Curvature along the null geodesics}
\label{appen:ScalarCurvature}

Singularities are boundary points of  spacetime where the
normal differentiability breaks down. 
If the energy density or the curvature invariant diverges at boundary
points,  the points are singularities.
In this appendix, we give the scalar curvature
$R^{\mu}_{\mu}$ in the HIN spacetime along the  radial null geodesics
terminating at the center $r=0$, and show that the center is singular
along the geodesics with $\alpha =7/9$ and $1$.

The scalar curvature $R^{\mu}_{\mu}$ of the HIN spacetime is given by 

\begin{equation}
R^{\mu}_{\mu} = \frac{8 \pi R^2}{R^2+16F^2} \, \epsilon
=  \frac{ \sqrt{2}}{(R-4m)\sqrt{m R}\sqrt{H}}.
\end{equation}
Along the geodesics $R = 2 x m^{1/3}$,  which are shown to be regular
null geodesics, $R^{\mu}_{\mu}$ is given by

\begin{equation}
R^{\mu} _{\mu} = \frac{3}{4x^3}+{O}(m^{2/3}),
\end{equation}
and it  is finite at $m=0$.
On the other hand, when we consider $R^{\mu}_{\mu}$ along the
geodesics, $R=2 x_0 m^{7/9}$ and 
$ R = 4m (1+y)$ of Eq. (\ref{eq:solution-y}),
the scalar curvatures are given by 

\begin{equation}
R^{\mu} _{\mu}= \frac{9}{28{x_{0}}^3 m^{4/3}}+ {O}(m^{-10/9}),
\end{equation}
and 
\begin{equation}
R^{\mu} _{\mu}= \frac{1}{64 m^{2}}+{  O}( m^{-5/2} y),
\end{equation}
respectively.
The scalar curvature diverges as  $m\to 0$ in these cases,  and
therefore the center is definitely singular.

\section{contraction mapping principle}
\label{appen:uniqueness}

We prove the existence of the  unique continuous solution of
Eq. (\ref{eq:RegularNulleq}). It is based on the discussion given by
Christodoulou for  the LTB model \cite{christodoulou1984}. 
Consider the differential equation obtained from
Eq. (\ref{eq:RegularNulleq}) by  replacing $\vartheta$ in $\Psi_{\mp}(\chi ,\vartheta)$ by a given continuous
function $\bar{\vartheta}>0$:

\begin{equation}
\frac{ d \vartheta}{d\chi }+\frac{6}{\chi }(\vartheta-\lambda _0)=\lambda _0
\Psi_{\mp}\left( \chi , \bar{{\vartheta}}(\chi) \right).
\label{eq:RegularNulleq-bar}
\end{equation}
Since we have chosen $\lambda = \lambda_0$,  $\Psi _{\mp}(\chi , \bar{\vartheta})$ is at least $C^1$
in the strip $ \chi \in [0 , \chi _1]$ and  $\bar{\vartheta} \in (0, \mu_1 ]$,  where we assume $\chi _1$ is sufficiently small.
The continuous solution of Eq. (\ref{eq:RegularNulleq-bar}) is only the
solution with $\vartheta(0)=\lambda _0(> 0)$.
This solution is given by

\begin{equation}
\vartheta(\chi ) = \lambda _0 \left( 1+\chi  \int ^1 _0 s^6 \Psi_{\mp}\left
( s \chi ,\bar{\vartheta} (s \chi) \right) ds \right).
\label{eq:solution-thetabar}
\end{equation}

We consider the nonlinear map $T_{\lambda _0}$ defined as
$\vartheta=T_{\lambda _0}(\bar{\vartheta})$. 
Since we are considering the function $\bar{\vartheta}$  
 with $\bar{\vartheta}(0)>0 $, we obtain $T_{\lambda
_0}(\bar{\vartheta})(0)=\lambda_0$ from expansions of
Eqs. (\ref{eq:JmpExpantion-z}) and (\ref{eq:sHExpantion-z}).
Thus the possible fixed point of $T_{\lambda_0}$ is only one which satisfies 
${\vartheta _{\rm{FP}}}(0)=\lambda_0$. To prove the existence of the
fixed point, we consider the set $V_{\mu}$ consisting of all
$\bar{\vartheta}$ such that for $\chi \in [0, \chi _2] $,

\begin{equation}
 \mu_{-} \leq  \bar{\vartheta}(\chi) \leq  \mu_{+} ,
\end{equation}
where the lower and the upper bound are defined by  $\mu_{-} \equiv
\lambda_0-\mu_2 $ and $\mu_{+} \equiv \lambda_0 + \mu_2$ for a
sufficiently small $\mu_2$.
From these bounds,  the nonlinear  map is also restricted in 

\begin{equation}
\mu_{-} \leq  T_{\lambda _0}(\bar{\vartheta})(\chi ) \leq \mu_{+} ,
\end{equation}
if we choose a $\chi$  such that 

\begin{equation}
 \chi \leq \chi_3 \equiv \frac{7 \mu_2 }{\lambda_0 \Delta_1} ,
\end{equation}
where

\begin{equation}
\Delta _1 = \sup_{0 \leq \chi \leq \chi_2 \, } \sup_{\mu_{-} \leq
\bar{\vartheta} \leq  \mu_{+}} |\Psi _{\mp}|.
\end{equation}
The map $T _{\lambda _0}$ sends $V _{\mu}$ into itself for all  $\chi
_4 \leq  \min \{ \chi _2, \chi_3 \}$.  Let

\begin{equation}
\Delta _2 = \sup _{0 \leq \chi  \leq \chi _2\,}
\sup_{\mu_{-} \leq
\bar{\vartheta} \leq  \mu_{+} } \biggl| \frac{\partial \Psi_{\mp}}{\partial
\bar{\vartheta}} \biggr|.
\end{equation}
Then we obtain from Eq. (\ref{eq:solution-thetabar}) for $\bar{\vartheta} _1,\,
\bar{\vartheta}_2 \in V_{\mu}$,

\begin{equation}
\| T_{\lambda _0}(\bar{\vartheta}_1)- T_{\lambda _0}(\bar{\vartheta}_2) \| 
\leq \frac{\chi _4 \lambda _0 \Delta_2 }{7} \| \bar{\vartheta}_1 -\bar{\vartheta}_2  \|,
\end{equation}
where  $\|$ $ \|$ denotes the supremum norm.
If we choose $\chi _4 < 7/(\lambda _0 \Delta_2) $, the map $T_{\lambda _0}$ 
is contractive in $V_{\mu}$. Hence  $T_{\lambda_0}$ has a unique
fixed point $\vartheta _{\rm{FP}} \in V_{\mu}$ which is given by 
\begin{equation}
\vartheta_{\rm{FP}}(\chi ) = T_{\lambda _0}(\vartheta_{\rm{FP}})(\chi ) 
= \lambda _0  \left( 1+\chi  \int ^1 _0 s^6
\Psi_{\mp} \left( s \chi ,\vartheta_{\rm{FP}}(s \chi) \right) ds \right).
\end{equation} 
Therefore we have the  unique continuous  solution with $\vartheta (0)=\lambda
_0$.

\section{derivation of eq. (\ref{EQ:CPNFORMALLYFLAT})}
\label{appen:Null}

In this appendix, we construct double null coordinates by using
Eq. (\ref{eq:solution-y}). We define double null coordinates $(u,v)$
as Eqs. (\ref{eq:defineu}) and (\ref{eq:definev}). 
From this definition,  $dm$ and $dR$ are given by

\begin{eqnarray}
dm  &=&  p_{u}du + p_{v}dv,
\\
dR  &=& 4dm +(q_u du + q_v dv),
\end{eqnarray}
where
\begin{eqnarray}
p_{u}  &\approx& -p_{v}\approx 2 (u-v),
\label{eq:pupv}
\\
q_u &\approx & - 4 \sqrt{2} y  \left( v+u -\sqrt{2} \beta (v-u)^{1/12}
+(\sqrt{2}+1)(v-u)+ \frac{\sqrt{2}}{5} \gamma \, (v-u)^{5/12} \right) du,
\\
q_v &\approx &   4 \sqrt{2} y  \left( v+u -\sqrt{2} \beta (v-u)^{1/12}
+(\sqrt{2}-1)(v-u)+ \frac{\sqrt{2}}{5} \gamma\, (v-u)^{5/12}
\right) dv .
\end{eqnarray}
The line element of Eq. (\ref{eq:ds})  is rewritten by the coordinate transformation from $(m,
R)$ to $(u,v)$

\begin{eqnarray}
ds^2 &=& U du^2+V dv^2 +W dudv + R^2 d\Omega ^2,
\label{eq:ds-null}
\end{eqnarray}
where the metric functions  $U,V$ and $ W$ are given by
\begin{eqnarray}
U &=& -(A+8B+16C) p_{u}^2 -C q_{u}^{2} - 2(B+4C) p_{u}q_{u},
\label{eq:metricU}
\nonumber
\\
V &=&  -(A+8B+16C) p_{v}^2 -Cq_v^2-2(B+4C)p_{v}q_v,
\label{eq:metricV}
\\
W &=& -2(A+8B+16C) p_{u}p_{v} -2C q_{u}q_{v} - 2(B+4C)(p_{u}q_{v}+p_{v}q_{u}).
\label{eq:metricW}
\nonumber
\end{eqnarray}
From Eq. (\ref{eq:pupv}), we obtain

\begin{equation}
  W \approx -U-V -C( q_u+ q_v)^2,
\end{equation}
and then Eq. (\ref{eq:ds-null}) becomes
\begin{equation}
ds^2 \approx -C(q_u+ q_v)^2 dudv+ U (du^2-dudv)+ V (dv^2-dudv)+R^2 
d\Omega ^2.
\end{equation}
If the condition
\begin{equation}
-C(q_u+ q_v)^2 \gg  U, V
\label{eq:nullcondition}
\end{equation} 
is satisfied, the double null coordinates would be constructed approximately.

We expand $\sqrt{H}$ of Eq. (\ref{eq:expantionH-atm0}) around $y=0$,
\begin{equation}
\sqrt{H}
= h(m)  +\sqrt{2} \left( \frac{8}{y} -4\, {\ln y} + 4 \,(3+\delta)+ y
+\frac{3}{4} y^2  + O(y^3) \right),
\label{eq:expantion-Hh(m)}
\end{equation}
where $h(m)$ is defined as 
\begin{equation}
h(m) = 4\sqrt{2} \left( \frac{\beta}{m^{1/3}} +\, \gamma \, m^{1/3}+{O}(m) \right).
\end{equation}
Inserting Eq. (\ref{eq:expantion-Hh(m)}) into $A$, $B$ and
$C$, metric functions $U$, $V$ and $W$ are calculated. 
As a result we obtain  
\begin{eqnarray}
\lim _{m \to 0} -\frac{1}{m}C(q_u+ q_v)^2 &=&   -512, 
\nonumber
\\
\lim _{m \to 0} \frac{U}{m}  &=& 0,
\\
\lim _{m \to 0} \frac{V}{m}  &=& 0, 
\nonumber
\end{eqnarray}
and it satisfies the condition of Eq. (\ref{eq:nullcondition}).
Hence the line element  around the center is given by

\begin{equation}
ds^2 \approx - 512(v-u)^2 \, dudv + R^2 d\Omega ^2.
\end{equation}



\end{document}